# The NIAID Discovery Portal: A Unified Search Engine for Infectious and Immune-Mediated Disease Datasets

#Ginger Tsueng[a], #Emily Bullen[a], Candice Czech[a], Dylan Welzel[a], Leandro Collares[a], Jason Lin[a], Everaldo Rodolpho[a], Zubair Qazi[a], Nichollette Acosta[a], Lisa M. Mayer[b], Sudha Venkatachari[c], Zorana Mitrović Vučičević[e], Poromendro N. Burman[e], Deepti Jain[e], Jack DiGiovanna[e], Maria Giovanni[d], Asiyah Lin[b]**, Wilbert Van Panhuis[b], Laura D. Hughes[a]*, #Andrew I. Su[a], #Chunlei Wu[a],

[a]The Scripps Research Institute, La Jolla, CA

[b]Office of Data Science and Emerging Technologies, National Institute of Allergy and Infectious Diseases, Rockville, MD

[c]National Cancer Institute, Rockville, MD

[d]National Institute of Allergy and Infectious Diseases, Rockville, MD

[e]Velsera, Charlestown, MA

Running head: "NIAID Discovery Portal: Unified Dataset Search"

#Address correspondence to Andrew I. Su, asu@scripps.edu; Chunlei Wu, cwu@scripps.edu; Ginger Tsueng, gtsueng@scripps.edu; Emily Bullen, ehaag@scripps.edu

*Present address: Laura D. Hughes, Genentech, San Francisco, CA 94080

**Present address: Asiyah Yu Lin, OntoData Research and Solutions LLC, Bethesda, MD 20817

## ABSTRACT

The NIAID Data Ecosystem Discovery Portal (https://data.niaid.nih.gov) provides a unified search interface for over 4 million datasets relevant to infectious and immune-mediated disease (IID) research. Integrating metadata from domain-specific and generalist repositories, the Portal enables researchers to identify and access datasets using user-friendly filters or advanced queries, without requiring technical expertise. The Portal supports discovery of a wide range of resources, including epidemiological, clinical, and multi-omic datasets, and is designed to accommodate exploratory browsing and precise searches. The Portal provides filters, prebuilt queries, and dataset collections to simplify the discovery process for users. The Portal additionally provides documentation and an API for programmatic access to harmonized metadata. By easing access barriers to important biomedical datasets, the NIAID Data Ecosystem Discovery Portal serves as an entry point for researchers working to understand, diagnose, or treat IID.

## IMPORTANCE

Valuable datasets are often overlooked because they are difficult to locate. The NIAID Data Ecosystem Discovery Portal fills this gap by providing a centralized, searchable interface that empowers users with varying levels of technical expertise to find and reuse data. By standardizing key metadata fields and harmonizing heterogeneous formats, the Portal improves data findability, accessibility, and reusability. This resource supports hypothesis generation, comparative analysis, and secondary use of public data by the IID research community, including those funded by NIAID. The Portal supports data sharing by standardizing metadata and linking to source repositories, and maximizes the impact of public investment in research data by supporting scientific advancement via secondary use.

---

## INTRODUCTION

The biomedical research community has entered an era in which open science and data sharing are increasingly regarded as essential to accelerating discovery, improving reproducibility, and ensuring the integrity of publicly funded research. These expectations have been formalized by the U.S. National Institutes of Health (NIH) through its Data Management and Sharing Policy, which went into effect in 2023 and requires researchers to plan for and make their data publicly available whenever possible (1). Similar mandates from scientific journals and funding agencies reflect a broader shift toward transparency and reuse in science.

While this emphasis on data sharing has led to a proliferation of public datasets (2), it has not eliminated the core challenges facing researchers who wish to reuse existing data. A critical and often overlooked barrier is the *discovery* of relevant datasets (3). Biomedical data are housed in hundreds of public repositories, both domain-specific and generalist, each with different metadata standards, access protocols, and search capabilities (4). As a result, finding datasets suitable for reuse often requires prior knowledge of where those datasets are hosted, what metadata they use, and how to navigate their interfaces. Even experienced data scientists frequently rely on ad hoc strategies, such as citation mining or keyword searching across multiple websites, to identify usable datasets (5).

These challenges are particularly acute in the field of **infectious and immune-mediated disease (IID)** research, which spans diverse experimental domains, from immunology and microbiology to clinical trials, imaging, and multi-omics. For example, a researcher studying host response to influenza may need to integrate transcriptomics data from GEO (6), immunophenotyping data from ImmPort (7), and viral sequences from GenBank (8). Yet each of these repositories uses its own metadata conventions, and none provides the means to search across them in a unified way. Inconsistent annotations for key properties such as species, pathogen, health condition, or assay type further impede cross-repository search, even when the data are technically accessible.

Poorly formatted metadata remains a major obstacle, limiting the usability of otherwise available data. There is growing recognition that dataset metadata is as crucial as the datasets themselves for ensuring that shared data can be found, interpreted, and reused. The FAIR data principles (Findable, Accessible, Interoperable, Reusable) have helped articulate this need by emphasizing not only data availability, but also machine-readable, semantically rich, and standardized metadata (9). Efforts such as [Schema.org](Schema.org) (10), Bioschemas (11), and the NIH Generalist Repository Ecosystem Initiative (GREI) (12) have begun to promote more consistent metadata practices. However, these initiatives rely heavily on repository participation and metadata provider compliance, something that is difficult to enforce and slow to propagate.

To address these problems, we developed the NIAID Data Ecosystem (NDE) Discovery Portal ([https://data.niaid.nih.gov](https://data.niaid.nih.gov)), a centralized interface that allows researchers to find datasets related to infectious and immune-mediated diseases regardless of where they are stored. Rather than attempting to centralize the data itself or enforce new metadata standards at the point of submission, our approach harmonizes metadata *post hoc* across multiple repositories using a custom schema based on existing standards. We then expose this harmonized metadata through a search engine tailored to the needs of IID researchers.

The rationale for this work is simple: shared data cannot fulfill its potential unless it can be readily found and reused. By focusing on post-hoc harmonization of repository metadata, a federated approach that respects the autonomy of participating repositories, and user-friendly search, the Discovery Portal enables researchers to identify relevant datasets for hypothesis generation, secondary analysis, and validation. In this resource report, we describe the development, design principles, and current capabilities of the NIAID Discovery Portal, with particular attention to the value it provides to end users in the biomedical research community.

---

## RESULTS

The NDE Discovery Portal was developed to serve as a unified entry point for finding publicly available datasets relevant to infectious and immune-mediated disease (IID) research.

Recognizing the diversity of experimental approaches and data types used in IID science, the Discovery Portal is designed to integrate metadata from a broad range of repositories spanning both domain-specific and generalist infrastructures. The result is a comprehensive, cross-repository index that allows researchers to identify relevant datasets without needing to know in advance which repository houses them or how that repository structures its metadata.

As of this writing, the Discovery Portal aggregates and harmonizes metadata from more than 4.3 million datasets across 42 repositories. These include prominent **domain-specific repositories** such as NCBI Sequence Read Archive (SRA) (13) for next-generation sequencing data, Gene Expression Omnibus (GEO) (6) for transcriptomics and functional genomics studies, ImmPort (7) for immunology-focused clinical and mechanistic studies, MassIVE (14) for mass spectrometry proteomics datasets, Qiita (15) for microbiome and metagenomic data, and COVID RADx Data Hub (16) for data from COVID-19 diagnostic and translational research. In addition, the Portal includes metadata from several **generalist repositories** supported by the NIH and other funders, including Figshare (17), Zenodo (18), Dryad (19), Harvard Dataverse (20), and Vivli (21). The home page includes a searchable table of all indexed repositories, annotated with data type, research domain, and access criteria (Table 1).

This breadth allows researchers to query within a single interface for datasets across a wide spectrum of data modalities, from viral genomes to host transcriptomics, clinical trial data, and proteomic assays. For each integrated repository, metadata are harvested and mapped to a harmonized schema that aligns with FAIR principles, enabling consistent filtering and search across otherwise incompatible data sources.

Central to the utility of the Discovery Portal is a search interface that enables researchers to rapidly query the entire metadata index using free text or keyword-based input (Figure 1). This search is complemented by an array of intuitive, prebuilt filters that reflect metadata elements of particular interest to IID researchers. These filters include host species (e.g., *Homo sapiens*, *Mus musculus*), pathogen or infectious agent (e.g., *Mycobacterium tuberculosis*, *SARS-CoV-2*), health condition (e.g., asthma, HIV infection), and measurement technique (e.g., RNA-seq, flow cytometry).

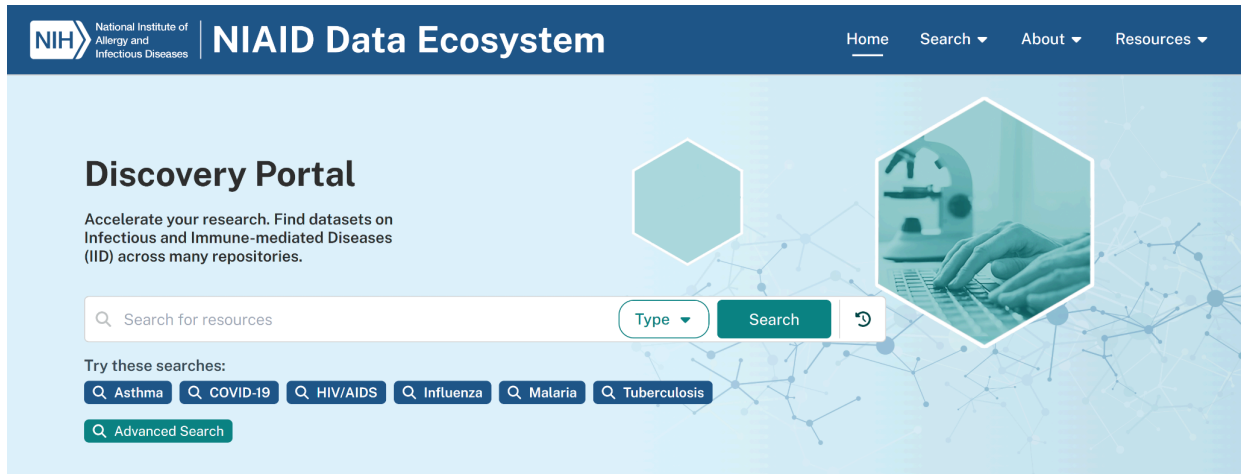

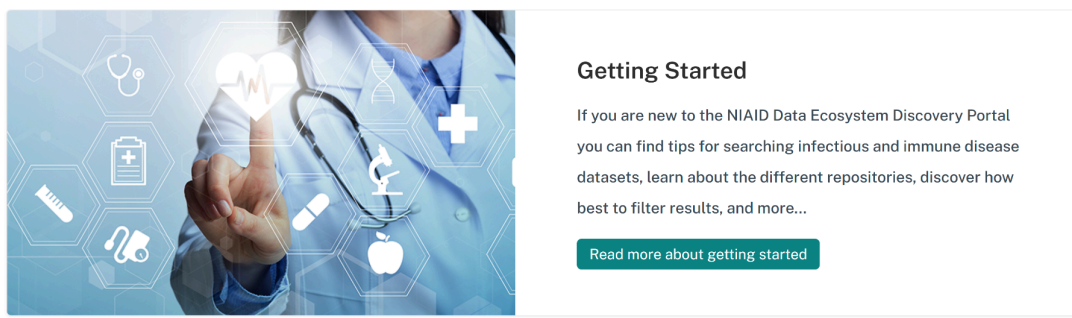

Figure 1. NIAID Data Ecosystem landing page and basic search interface.

These filters allow users to quickly refine results and identify datasets that are most relevant to their research needs. For example, a virologist can search for "Zika virus" and then restrict results to "Human" datasets measured via "Proteomics" in just a few clicks. The Portal also displays detailed metadata for each dataset and provides direct links to the originating repository for data access. By leveraging the Portal's harmonized metadata, researchers can refine their search, as well as perform downstream analyses across the datasets they identify.

For users with more complex information needs, the Portal also includes an advanced search interface that supports field-specific queries across nearly 50 metadata properties. This functionality is particularly useful for computational biologists, data managers, and tool developers who need to identify datasets with highly specific attributes, e.g., studies funded by a particular NIH grant, or datasets involving *Plasmodium falciparum* infection in nonhuman primates using ELISA assays. While the mechanics of query construction are hidden from users in the Basic Search interface, the Advanced Search option provides the flexibility to support compound filters, precise Boolean logic, and reproducible queries.

## DISCUSSION

The design of the NIAID Discovery Portal emphasized both the breadth of dataset coverage and the accessibility of the search interface with the intent of providing a powerful tool for facilitating data reuse across the IID research landscape. This goal of promoting data reuse is shared across the NIH, with similar institute-specific initiatives in the NIDDK Central Repository (https://repository.niddk.nih.gov/), the Index of NCI Studies (https://studycatalog.cancer.gov/), and the NIMH Data Archive (https://nda.nih.gov/). While the NLM also offers a broadly-scoped effort in the NLM Data Catalog (https://datasetcatalog.nlm.nih.gov/), these domain-specific efforts allow for the development of more focused metadata schemas (e.g., the tracking of `species` and `infectiousAgent` in the the NIAID Discovery Portal) that are specifically relevant to certain research communities.

While the NDE Discovery Portal improves the findability and accessibility of IID datasets, several limitations remain that present opportunities for future development.

A primary challenge is the incompleteness and inconsistency of metadata across repositories. Many records lack key fields, such as `species`, `infectiousAgent`, or `healthCondition`, which are essential for effective filtering. Although metadata augmentation strategies have substantially improved coverage, they cannot fully compensate for missing or ambiguous source metadata. Furthermore, variations in term formatting (e.g., capitalization, punctuation, and hyphenation) can lead to duplicate or fragmented filter values, especially in free-text fields. Our standardization pipeline attempts to reconcile these differences, but consistent use of community identifiers would result in a more accurate result.

Another limitation involves semantic inconsistencies and limited context in generalist repositories. These platforms often include minimal required metadata, making it difficult to infer experimental detail or biological relevance that is necessary to enable domain-specific reuse. While additional metadata can sometimes be extracted from linked publications or supplementary materials, persistent identifiers (e.g., DOIs, PubMed IDs) are often not captured in a systematic and consistent manner.

From a technical standpoint, the Discovery Portal relies on a schema translation and indexing model that supports flexible integration but requires ongoing maintenance. Repository APIs, data formats, and metadata standards continue to evolve, necessitating regular updates to ingestion pipelines and harmonization logic. Maintaining a scalable, performant system while integrating new sources will remain a key operational challenge.

Looking forward, we plan to expand the scope and functionality of the Portal along several axes:

- **Incorporating additional resource types**, including computational tools, software pipelines, and models linked to datasets, to support reproducibility and reuse.
- **Improving entity resolution** through integration of cross-references (e.g., UMLS, MeSH, UniProt) and persistent identifiers, especially for diseases and organisms.
- **Supporting multilingual and lay-access metadata enhancements**, particularly for datasets relevant to global infectious disease surveillance and public health.

- **Exploring natural language and AI-assisted search interfaces**, including the integration of large language models (LLMs), to support more intuitive querying by non-specialist users.
- **Enhancing community engagement**, including support for user-contributed metadata curation, tagging, and annotation.

Ultimately, the continued value of the Discovery Portal will depend on its ability to evolve alongside the data practices and priorities of the IID research community. By identifying limitations and addressing them transparently, we aim to support a more FAIR, interconnected, and discoverable ecosystem of biomedical data.

## MATERIALS AND METHODS

### Metadata Schema and Mapping

To support consistent and effective search across diverse data repositories, the NDE Discovery Portal relies on a unified metadata schema that harmonizes descriptive information about datasets from disparate sources. The goal of this schema is to balance breadth and specificity, capturing a wide range of dataset types across IID research while maintaining semantic consistency that supports meaningful filtering, indexing, and interoperability.

The NDE dataset metadata schema was initially developed by the NIAID Systems Biology Data Dissemination Working Group (22). The approach is based on the Schema.org Dataset class (10), a widely adopted standard in biomedical and general web search contexts. This foundation was extended with additional properties relevant to IID research, informed by user interviews, repository landscape analysis, and FAIR data principles.

The Dataset schema consists of a core set of **required, recommended, and optional fields**, including:

- **Required properties**:
    - `name` (dataset title)
    - `description`
    - `identifier` (e.g., DOI or accession number)
    - `url` (link to the data source)
    - `author` – author or creator of the data
    - `funding` – grant or program that supported the data generation
    - `measurementTechnique` – methods used (e.g., RNA-seq, flow cytometry)
    - `includedInDataCatalog` – original repository
    - `distribution` – data download information

- **Recommended properties**:
    - `healthCondition` – mapped to disease or clinical indication

- `infectiousAgent` – pathogen or organism of study
- `species` – host or model organism
- `variableMeasured` – key variables or outputs
- `keywords` – key terms helpful for search
- `doi` – a digital object identifier for the dataset itself (if available)
- `temporalCoverage`, `spatialCoverage` – for epidemiological or longitudinal data
- `conditionsOfAccess`, `license`, `usageInfo` – licensing or access terms
- `sdPublisher` – the original publisher of the structured metadata if available
- `dateCreated`, `dateModified`, `datePublished` – date information
- `citation`, `isBasedOn`, `citedBy` – citation and reference information
- **Optional properties**:
  - `hasPart, isPartOf, isRelatedTo, isSimilarTo, sameAs, isBasisFor` – information on how the Dataset relates to other types of CreativeWork
  - `nctid` – [ClinicalTrials.gov](ClinicalTrials.gov) identifier if available
  - `version` – version information
  - `abstract` – used only when a source has both a summary/abstract and description information
  - `isAccessibleForFree` – boolean indicating whether or not there is a cost to access
  - `sourceOrganization` – a project, program, or organization for which the data was generated
  - Any other mappable [schema.org](schema.org) Dataset property that can be mapped to a property used by a resource

Wherever possible, metadata values are aligned with standard biomedical ontologies. For example, `species` and `infectiousAgent` values are mapped to NCBI Taxonomy IDs (23), `healthCondition` values are mapped Mondo Disease Ontology (MONDO) (24), Human Phenotype Ontology (HPO) (25), and Disease Ontology (DOID) (26), `topicCategory` values are mapped to EDAM (27), and `funding.identifier` values are mapped to NIH grant numbers.

To facilitate metadata integration, each target repository is assessed for compatibility and relevance, with particular focus on repositories commonly cited in IID research or supported by NIAID funding. For each repository, a custom parser and transformation pipeline was implemented to extract, normalize, and map native metadata fields to the NDE schema. For instance:

- In **NCBI SRA**, elements such as study title, organism, and sequencing platform are extracted from XML records and mapped to `name`, `species`, and `measurementTechnique`, respectively.
- In **GEO**, metadata fields describing array or RNA-seq experiments are mapped to `variableMeasured`, `measurementTechnique`, and `healthCondition`, often leveraging accompanying PubMed references for context.
- In **ImmPort**, metadata about condition, species, and assay type can be directly mapped to `healthCondition`, `species`, and `measurementTechnique`.
- In **AccessClinicalData@NIAID**, clinical trial metadata such as titles, identifiers, summaries, and conditions, are mapped to `name`, `identifier`, `description`, and `healthCondition`.
- In **VEuPathDB**, scientific context including organism, gene counts, and measurement types are mapped to `species`, `variableMeasured`, and `measurementTechnique`, respectively.
- For **generalist repositories** (e.g., Zenodo, Figshare), where metadata quality is more variable, the mapping pipeline prioritizes extraction of the most common standardized fields (e.g., `name`, `description`, `doi`) and infers additional properties when possible using linked resources such as citations.

Each repository-specific parser converts the native metadata into a JSON-LD document conforming to the NDE schema. These documents are then indexed in Elasticsearch (28) and exposed via the Discovery Portal's search interfaces and public API.

In designing the NDE schema, we aligned our approach with broader community standards to ensure compatibility and interoperability. These include the Bioschemas initiative (11), which extends Schema.org to meet life science data needs, and GREI (12), which aims to improve metadata harmonization across NIH-supported generalist repositories. By drawing on these efforts, the NDE schema ensures alignment with emerging best practices and supports future interoperability with external search engines, FAIR data catalogs, and semantic web tools.

This schema-based harmonization strategy enables datasets from dozens of heterogeneous repositories to be searched, compared, and filtered in a coherent and consistent manner, without requiring repositories to adopt a new internal data model. Importantly, the use of open standards and publicly documented schema definitions also ensures that the Discovery Portal can support interoperability with external tools and data discovery platforms.

## Metadata Augmentation

Despite the adoption of a harmonized metadata schema, many records aggregated by the NIAID Discovery Portal lacked key metadata elements that are essential for meaningful filtering and dataset discovery, particularly for fields like `species`, `infectiousAgent`, `healthCondition`, and `funding`. These gaps reflect the heterogeneity in metadata practices across repositories, particularly among generalist platforms and older records. To address this limitation and improve the user experience, we implemented a multi-pronged metadata augmentation strategy to enrich and standardize metadata values post-ingestion.

For structured fields such as `species` and `infectiousAgent`, we used Text2Term (29), a local ontology mapping tool, to match free-text values to standardized terms from the NCBI Taxonomy (23). This process included extracting scientific, common, and alternative names as well as taxonomic identifiers, enabling consistent filter behavior and synonym resolution in the Discovery Portal interface. Since host species and pathogenic agents are both often provided under a single "organism" label in source repositories, we applied a taxonomy-based heuristic. This heuristic leveraged lineage information from the NCBI Taxonomy tree to classify organisms as potential hosts (e.g., vertebrates, arthropods, plants) or likely pathogens (e.g., bacteria, viruses, protozoa). Default assignments were manually reviewed for the most frequently occurring terms and iteratively refined based on user feedback.

For the `healthCondition` field, we implemented a hierarchical ontology mapping workflow using the NCATS Translator Knowledge Provider (KP) APIs (30). These services allowed us to normalize disease-related terms to multiple interoperable ontologies, following a defined priority order: we first attempted to map to the Mondo Disease Ontology (MONDO) (24), then to the Human Phenotype Ontology (HPO) (25) if no MONDO match was found, followed by the Disease Ontology (DOID) (26), and finally the NCI Thesaurus (NCIT) (31).

This approach ensured broad coverage and alignment with downstream analysis tools and data standards. Each mapped term was stored along with its identifier, preferred label, and known synonyms.

**Citation-Linked Metadata Augmentation**

When a dataset was linked to a PubMed-indexed publication, we used the citation as an external source of metadata to enhance the associated record. We extracted information from the publication, including diseases, organisms, and funding sources, using a combination of PubTator annotations (32) and custom filtering logic to restrict terms to those appearing in the dataset's `name` or `description`. This ensured that only highly relevant terms were used for augmentation.

This citation-based strategy significantly improved metadata completeness, particularly for the `healthCondition` and `funding` fields. In many cases, metadata extracted from the publication was the only structured information available to identify the biological context of a dataset.

**Text Mining for Unlinked Records**

For datasets lacking associated publications or standardized terms, we used the EXTRACT tool to identify biological concepts directly from free-text fields such as `description`. While this method is inherently noisier, it provided valuable metadata for records that would otherwise be unannotated. We established internal quality controls and a review protocol for correcting or suppressing incorrectly inferred terms.

Based on the metadata augmentation pipelines above, we were able to substantially increase the completeness in our metadata catalog (Figure 2).

| # of records with | Ingest only | Standardization and delineation | Citation-based augmentation | EXTRACT augmentation |
|---|---|---|---|---|
| species | 163,475 | 200,837 | 201,225 | 1,084,271 |
| infectiousAgent | 386 | 5,011 | 6,097 | 441,473 |
| healthCondition | 7,188 | 7,188 | 115,649 | 983,602 |
| funding ID | 28,187 | 28,187 | 162,580 | 664,772* |
| Total records** | 2,828,146 | 2,833,170 | 2788765 | 3,355,155 |
| | | | | |
| # of different | | | | |
| species | 1,195 | 342 | 450 | 35,205 |
| infectiousAgent | 41 | 894 | 1,015 | 43,090 |
| healthCondition | 571 | 571 | 3,296 | 8,653 |

*Values include parser improvements & changes to repository exports which previously did not provide structured funding data
**Total records available during snapshot. This includes increases due to additional repositories and repository updates

Figure 2. Overview of metadata improvements.

We used PubTator (32) for free-text augmentation when PubMed IDs were available and EXTRACT when they were not. As both tools were evaluated by the BioCreative challenge (33), we used these benchmarks and focused our internal quality control efforts on iteratively removing terms that EXTRACT frequently mislabels. These corrections are maintained in a public corrections repository. For structured field standardization, we benchmarked our use of the ontology mapping tool, Text2Term (29), against PubTator and found only a 0.038% difference in mapping outcomes.

**Augmentation of Topic Categories**

To improve domain-level filtering and support topic-based navigation, we added the `topicCategory` field to each record using the EDAM Topics ontology (27). An initial set of ~380 records across all repositories were manually annotated and used as a benchmark to evaluate classification performance by a large language model (ChatGPT) (34). The model's predictions were scored against human raters using an adjusted agreement metric that accounted for EDAM's hierarchical structure. Based on its consistent performance, the model was used to annotate over 1.8 million records with EDAM topic classifications.

By augmenting and standardizing metadata across key fields, we significantly improved the completeness, consistency, and usability of records in the Discovery Portal. These efforts not only enhance the precision and relevance of search results, but also increase the visibility of datasets that might otherwise remain hidden due to poor metadata quality. Together with schema harmonization, metadata augmentation is a critical component of our strategy to make IID-related datasets more FAIR and actionable for the research community (Figure 3).

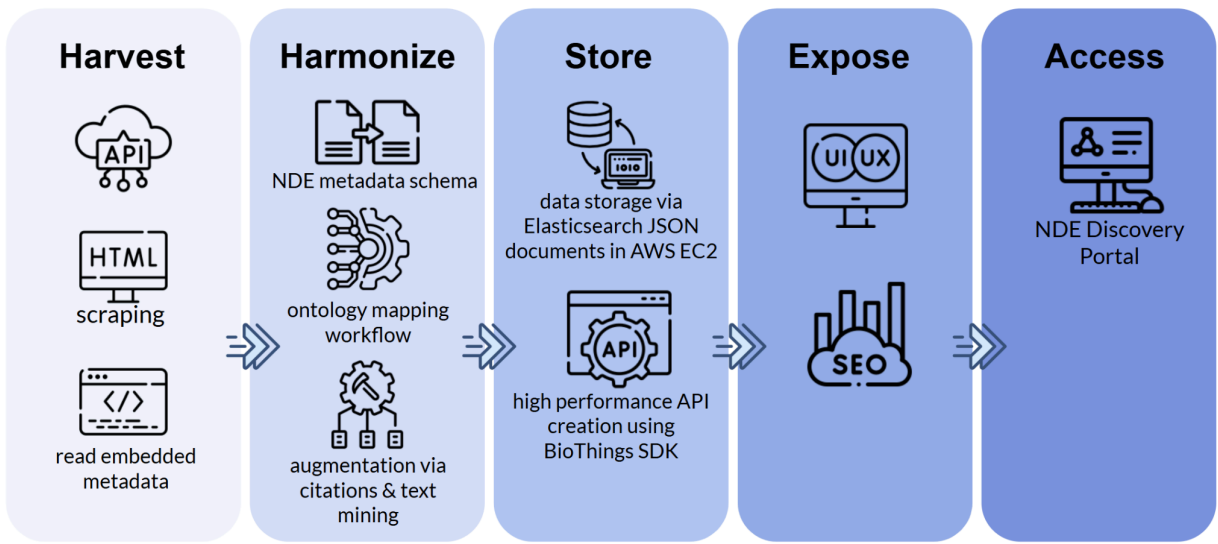

Figure 3. Overview of the NIAID Data Ecosystem metadata integration pipeline.

## Outreach and Usage Metrics

To promote awareness and adoption of the NIAID Discovery Portal, we implemented a multifaceted outreach strategy targeting the IID research community. This included presentations at NIH workshops and scientific conferences, targeted email campaigns to NIAID-funded investigators, social media engagement, and content development for blogs and data science forums. In parallel, we optimized the Portal for discoverability via search engine optimization (SEO) strategies (35), including enhanced metadata tagging, improved mobile performance, and backlink generation from trusted scientific and institutional websites.

These efforts led to a steady increase in Portal engagement. Since its launch, the Discovery Portal has attracted over 100,000 unique users, with more than 180,000 pageviews and growing. Usage analytics show that both the basic search and advanced search functionalities are regularly used, with a notable proportion of sessions involving metadata filtering by species, pathogen, or health condition. Peaks in user activity have consistently corresponded with outreach events, underscoring the importance of sustained community engagement.

The Portal currently receives over 10,000 monthly users, reflecting its growing role as a central resource for data discovery in the IID research community. In addition, we have also collaborated with several IID domain specific research programs, such as the Systems Biology Consortium for Infectious Diseases (36), Centers for Research in Emerging Infectious Diseases (CREID) Network (37), and Research and Development of Vaccines and Monoclonal Antibodies for Pandemic Preparedness (REVAMPP) (38), to make their datasets more discoverable and accessible by researchers.

## Data and Code Availability

- Portal: https://data.niaid.nih.gov
- API: https://api.data.niaid.nih.gov
- Metadata schema: https://discovery.biothings.io/ns/nde
- Source code (planned for public release via GitHub)

## ACKNOWLEDGEMENTS

We gratefully acknowledge the strategic guidance and feedback provided by Meghan Hartwick and Reed S. Shabman in the Office of Data Science and Emerging Technologies (ODSET) at the National Institute of Allergy and Infectious Diseases (NIAID). This material is based upon work supported by the Frederick National Laboratory for Cancer Research operated by Leidos Biomedical Research, Inc under contract number **75N91020F00022**. Any opinions, findings, and conclusions or recommendations expressed in this material are those of the author(s) and do not necessarily reflect the views of the Frederick National Laboratory for Cancer Research, Leidos Biomedical Research, Inc., or NIAID.

Furthermore, the development and ongoing operation of the NIAID Data Ecosystem involve the support of the numerous data repositories who host the data and provide metadata. As described, the list of contributing repositories is continuously expanding; a current list of sources is available at https://data.niaid.nih.gov/sources.

## AUTHORS' CONTRIBUTIONS

EB conducted the user studies to inform the design, wrote the documentation pages, engaged in outreach, curated metadata for metadata augmentation. CC designed the front-end architecture for the site. CC and LC conducted user studies, designed and implemented the user interface. EB and CC implemented SEO to improve visitation to the site. DW, JL, and NA wrote the parsers for metadata ingestion, with JL and DW maintaining the parsers and metadata updates. DW, JL and ZQ augmented the metadata. DW, JL, and ER designed and maintained the API and back-end infrastructure. GT designed the schemas, mapped resources to the schemas, engaged in outreach, identified the methodologies for metadata augmentation and conducted analysis on the quality of the results, and curated metadata for Resource integration. GT managed the project with guidance from SV, LDH, AIS, and CW. LMM, MG, AYL, and WVP provided strategic guidance for the project. LMM and WVP provided design ideas, prioritization, and connected the team to stakeholders for feedback, improvement suggestions, and community engagement. LDH conceived and managed the first phase of the project. JD, DJ, ZMV and PNB contributed to the early design and architecture. AIS and CW supervised the project. AIS and EB wrote the paper with suggestions from GT and LMM.

assessment of information extraction for biology. BMC Bioinformatics 6.

34. OpenAI. 2023. ChatGPT. ChatGPT (Mar 14 version) [Large language model]. https://chatgpt.com/. Retrieved 17 June 2025.

35. Google Search Central. 2025. SEO Starter Guide: The Basics. Google for Developers. https://developers.google.com/search/docs/fundamentals/seo-starter-guide. Retrieved 9 July 2025.

36. National Institute of Allergy and Infectious Diseases. 2008. Systems Biology Consortium for Infectious Diseases. NIAID. https://www.niaid.nih.gov/research/systems-biology-consortium. Retrieved 9 July 2025.

37. National Institute of Allergy and Infectious Diseases. 2020. CREID Network. Centers for Research in Emerging Infectious Diseases. https://creid-network.org. Retrieved 22 June 2025.

38. RTI International. 2024. ReVAMPP. Research and Development of Vaccines and Monoclonal Antibodies. https://revampp.org/. Retrieved 22 June 2025.

**Table 1**

*Dataset repositories currently included in the NIAID Data Ecosystem*

| Name | Type | Research Domain | Access |
| --- | --- | --- | --- |
| AccessClinicalData@NIAID | Dataset Repository | IID | Varied Access |
| AmoebaDB | Dataset Repository | IID | Registered Access |
| ClinEpiDB | Dataset Repository | IID | Varied Access |
| COVID RADx Data Hub | Dataset Repository | IID | Unknown Access |
| CryptoDB | Dataset Repository | IID | Registered Access |
| Data Discovery Engine | Dataset Repository | Generalist | Varied Access |
| Database of Genotypes and Phenotypes (dbGaP) | Dataset Repository | Generalist | Controlled Access |
| Dryad Digital Repository | Dataset Repository | Generalist | Open Access |
| Figshare | Dataset Repository | Generalist | Unknown Access |
| Flow Repository | Dataset Repository | Generalist | Unknown Access |
| FungiDB | Dataset Repository | IID | Registered Access |
| GiardiaDB | Dataset Repository | IID | Registered Access |
| Harvard Dataverse | Dataset Repository | Generalist | Varied Access |
| HostDB | Dataset Repository | IID | Registered Access |
| HuBMAP | Dataset Repository | Generalist | Varied Access |
| Human Cell Atlas | Dataset Repository | Generalist | Varied Access |
| ImmPort | Dataset Repository | IID | Registered Access |
| ImmuneSpace | Dataset Repository | IID | Registered Access |
| LINCS | Dataset Repository | Generalist | Unknown Access |
| MalariaGEN | Dataset Repository | IID | Varied Access |
| MassIVE | Dataset Repository | Generalist | Open Access |
| Mendeley Data | Dataset Repository | Generalist | Varied Access |
| MicrobiomeDB | Dataset Repository | IID | Open Access |
| MicrosporidiaDB | Dataset Repository | IID | Registered Access |
| NCBI BioProject | Dataset Repository | Generalist | Open Access |
| NCBI GEO | Dataset Repository | Generalist | Unknown Access |
| NCBI SRA | Dataset Repository | Generalist | Varied Access |
| NICHD Data and Specimen Hub (DASH) | Dataset Repository | Generalist | Controlled Access |
| OmicsDI | Dataset Repository | Generalist | Unknown Access |
| PiroplasmaDB | Dataset Repository | IID | Registered Access |
| PlasmoDB | Dataset Repository | IID | Registered Access |
| Qiita | Dataset Repository | IID | Registered |
| ReframeDB | Dataset Repository | IID | Controlled Access |
| The Network Data Exchange (NDEx) | Dataset Repository | Generalist | Open Access |
| ToxoDB | Dataset Repository | IID | Registered Access |
| TrichDB | Dataset Repository | IID | Registered Access |
| TriTrypDB | Dataset Repository | IID | Registered |
| VDJServer | Dataset Repository | - | Open Access |
| VectorBase | Dataset Repository | IID | Registered Access |
| VEuPath Collections | Dataset Repository | IID | Registered Access |
| VEuPathDB | Dataset Repository | IID | Registered Access |
| Vivli | Dataset Repository | Generalist | Controlled Access |
| Zenodo | Dataset Repository | Generalist | Varied Access |